\documentclass{article}
\usepackage{letterspace}
\usepackage[T1]{fontenc}
\usepackage[utf8]{inputenc}
\usepackage{ismir}

\def\authorname{Y.-C. Cheng, J.-T. Wu, B. Chen, Y.-T. Yeh, Y.-H. Chen, and Y.-H. Yang}

\usepackage{amsmath,amssymb,amsfonts}
\usepackage{url}
\usepackage[bookmarks=false,hidelinks,pdfauthor={\authorname},pdfsubject={\pdfsubject}]{hyperref}

\usepackage{graphicx}
\usepackage{booktabs}
\usepackage{array}
\usepackage{multirow}

\usepackage{tikz}
\usetikzlibrary{arrows.meta,positioning,fit,backgrounds,calc,decorations.pathreplacing}

\usepackage{needspace}
\usepackage{enumitem}
\setlist{nosep,leftmargin=*}
\usepackage{xcolor}
\usepackage{float}

\newcommand{\stemfx}{\textsc{StemFX}}
\newcommand{\bsfilm}{BSFiLM}
\newcommand{\sepaug}{Sep-Aug}
\newcommand{\multiafx}{MultiAFx}
\newcommand{\eg}{e.g.}

\newcommand{\etal}{et~al.}

\title{StemFX: Learning Mixing Style Representations via Autoregressive FX Chain Prediction on Source-Separated Stems}
\multauthor
  {Yuan-Chiao Cheng$^1$ \hspace{0.7cm} Jui-Te Wu$^1$ \hspace{0.7cm} Brian Chen$^1$}
  {{\bf Yen-Tung Yeh$^2$ \hspace{0.7cm} Yu-Hua Chen$^2$ \hspace{0.7cm} Yi-Hsuan Yang$^3$}\\
  $^1$ Independent Researcher \hspace{0.5cm} $^2$ National Taiwan University, Taiwan\\
  $^3$ AI-CoRE, National Taiwan University, Taiwan\\
  {\tt\footnotesize im31132@gmail.com}
  }

\begin{document}
\maketitle

\pagestyle{plain}

\begin{abstract}
Audio mixing style encompasses the artistic and technical decisions a mix engineer makes, including level balancing, spatialization, and the choice, ordering, and parameterization of audio effects (FX) on each stem.
FX chains are a key determinant of this style, yet existing approaches to modeling them remain limited. Some operate on stereo mixtures without explicit per-stem FX chain modeling, others fix the number or type of effects per track, and many require differentiable effect implementations or scarce multitrack datasets.
We present \stemfx{}, a framework that learns mixing style representations by autoregressively predicting variable-length FX chains on source-separated stems.
A Transformer decoder predicts tokenized FX chains autoregressively, while a band-split multi-band CNN encoder with FiLM conditioning captures per-stem spectral structure.
To enable large-scale paired training, we extract pseudo-stems from about 105K songs via source separation and augment them using \multiafx{}, a toolkit unifying 85 audio effects from 7 Python libraries.
Evaluated on mixing style retrieval, \stemfx{} outperforms all baseline models across all tested chain lengths.
On paired mixing style transfer, \stemfx{} achieves the best spectral fidelity and the highest listener preference, over 4000 times faster than iterative optimization.
\end{abstract}

\section{Introduction}\label{sec:intro}

A mixing engineer applies a distinct chain of audio effects to each instrument track: an equalizer to brighten the vocals, a compressor to tighten the drums, reverb to place the guitar in a room.
The collective choice of effects, their ordering, and their parameter settings across all stems defines the \emph{mixing style}.
An FX chain can be thought of as a sentence: just as language derives expressiveness from an open vocabulary and flexible length, a faithful representation of mixing style should support variable-length chains with arbitrary effects.
Learning such representations would improve mixing style recognition, retrieval, and transfer.

Yet prior approaches severely constrain this vocabulary.
Contrastive encoders such as FxEncoder~\cite{koo2023fxencoder} require large batch sizes to provide sufficient negative pairs, increasing GPU memory cost.
Differentiable mixing approaches~\cite{vanka2024diffmst,steinmetz2021dmcnae,steinmetz2022deepafxst,steinmetz2023dasp} fix the number and type of effects per track and require differentiable implementations, restricting the set of usable effects.
ST-ITO~\cite{steinmetz2024stito} can optimize effect chains unseen during training, but only at inference time, costing thousands of forward passes and, in our experiments, over 1000 seconds per example; its AFx-Rep encoder learns effect representations through a classification pretext task over a fixed inventory of effect presets.
Most methods also require paired dry/wet multitrack stems, limiting them to small datasets: MUSDB18~\cite{rafii2017musdb} with 150 tracks, MedleyDB~\cite{bittner2014medleydb} with about 200, and MoisesDB~\cite{pereira2023moisesdb} with 240.

We propose \stemfx{}, which treats FX chain prediction as a sequence generation problem.
By predicting tokenized FX chains autoregressively on source-separated stems, \stemfx{} removes the constraints of fixed chain lengths and effect ordering.
Unlike DDSP methods~\cite{steinmetz2022deepafxst,steinmetz2023dasp}, it needs no differentiable implementations, predicting discrete token sequences rather than optimizing audio-domain losses; unlike methods that parameterize a fixed set of effect presets~\cite{vanka2024diffmst}, it predicts the effect structure and the chain length themselves.
The output is human-readable and interpretable, \eg{}, ``\texttt{sox\_compand $\to$ threshold=$-$20\,dB, ratio=4{:}1 $\to$ sox\_reverb $\to$ decay=1.5\,s},'' allowing engineers to inspect, modify, and execute the predicted chain.
Both the \stemfx{} codebase, with source code and trained model weights, and the \multiafx{} library are released as open source.\footnote{\url{https://github.com/barry-mir/stemfx} and \url{https://github.com/barry-mir/multiafx}}

Our contributions are as follows:
\begin{enumerate}
\item \textbf{\stemfx{} Framework.}
To the best of our knowledge, the first framework to jointly train an audio encoder and decoder end-to-end for learning mixing style representations through autoregressive FX chain generation.
This differs from LLM-based tool-calling approaches~\cite{doh2025llm2fx}, which keep the audio encoder frozen and fine-tune only the adapter and LLM via LoRA, so their generation objective never shapes the audio encoder.
\item \textbf{\bsfilm{} Encoder.}
A band-split multi-band CNN encoder~\cite{kim2021multiband} with Feature-wise Linear Modulation (FiLM) conditioning~\cite{perez2018film} from hand-crafted mixing features, bringing band-split processing to mixing style representation learning.
\item \textbf{\sepaug{} Pipeline and \multiafx{} Toolkit.}
Source separation applied to a large mixture dataset creates pseudo-stems, which are augmented with random chains drawn from \multiafx{}, our pip-installable wrapper over 85 effects from 7 Python libraries. Together they yield paired training data at about 105K-song scale, orders of magnitude larger than existing methods. See Section~\ref{sec:sepaug} for details.
\end{enumerate}

\section{Related Work}\label{sec:related}

\subsection{Automatic Mixing and Style Transfer}

Automatic mixing systems transfer the sonic characteristics of a reference mix to new material, through either representation learning or direct parameter prediction.
FxEncoder~\cite{koo2023fxencoder} disentangles mixing style from content via contrastive learning on multitrack audio, its successor Fx-Encoder++~\cite{yeh2025fxencoder} refines the architecture, and MEGAMI~\cite{koo2025megami} builds on these representations for conditional diffusion-based mixing.
Differentiable console methods such as Diff-MST~\cite{vanka2024diffmst} and the work of Steinmetz~\etal{}~\cite{steinmetz2021dmcnae} backpropagate through fixed effect chains, while ST-ITO~\cite{steinmetz2024stito} optimizes effect parameters at inference time.
LLM2Fx-Tools~\cite{doh2025llm2fx} is closest to our formulation, using autoregressive sequence modeling to generate the FX chain, but it keeps the audio encoder frozen and fine-tunes only a language model adapter via LoRA, so the generation objective does not improve the audio representations.
Collectively, these methods either omit explicit per-stem FX chain prediction, fix the effect structure, or rely on inference-time optimization. \stemfx{} sidesteps these constraints by processing separated stems at scale and jointly training encoder and decoder for variable-length FX chain prediction.

\subsection{Audio Representation Learning and Band-Split Architectures}

Self-supervised and contrastive audio representations such as CLAP~\cite{elizalde2023clap}, COLA~\cite{saeed2021cola}, CLMR~\cite{spijkervet2021clmr}, MERT~\cite{li2024mert}, Music2Latent~\cite{pasini2024music2latent}, and HTS-AT~\cite{chen2022htsat} target content-level tasks and do not capture fine-grained mixing characteristics.
Band-split architectures are effective for source separation~\cite{luo2023bsrnn,li2024scnet,lu2023bsroformer,wang2023melband} and multi-band CNNs for acoustic classification~\cite{kim2021multiband}, but have not been applied to mixing style representation.
We adapt this paradigm, using per-band CNNs to capture frequency-specific characteristics such as EQ curves and compression.

\begin{figure*}[t]
\centering
\resizebox{0.95\textwidth}{!}{%
\input{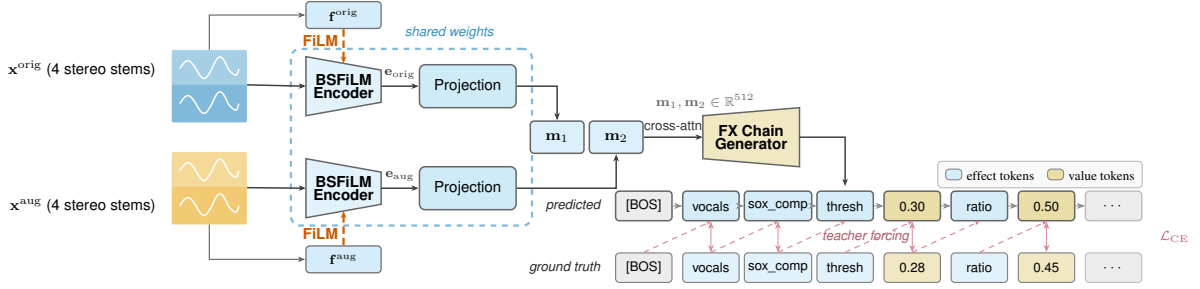}
}
\caption{\stemfx{} system overview.
Original and target stems pass through the shared \bsfilm{} Encoder to produce embeddings, projected into conditioning vectors $\mathbf{m}_1, \mathbf{m}_2$ that the Transformer decoder attends to via cross-attention.
The FX Chain Generator is trained with teacher forcing and cross-entropy loss against the next ground-truth token.}
\label{fig:system}
\end{figure*}

\section{Method}\label{sec:method}

\subsection{Framework Overview and Problem Formulation}\label{sec:problem}

Figure~\ref{fig:system} illustrates the \stemfx{} architecture.
The framework jointly trains a \bsfilm{} Encoder and a Transformer-based FX Chain Generator end-to-end. The encoder maps both the original and augmented stems into conditioning vectors that summarize the mixing style difference, and the generator autoregressively predicts the tokenized FX chain that was applied.
By treating FX chain prediction as a sequence generation problem, \stemfx{} supports variable-length chains with arbitrary effects.

We formalize the task as follows.
Let $\mathbf{x} \in \mathbb{R}^{8 \times T}$ denote a set of four source-separated stereo stems (vocals, bass, drums, other) at 44.1\,kHz.
Given an original input $\mathbf{x}^{\mathrm{orig}}$ and a target $\mathbf{x}^{\mathrm{aug}}$ augmented with a known FX chain, the goal is to predict the per-stem FX chain $\mathcal{F}$ that transforms $\mathbf{x}^{\mathrm{orig}}$ to match $\mathbf{x}^{\mathrm{aug}}$.
Each stem's chain is an ordered list of effect--parameter pairs:
\begin{equation}\label{eq:fx_chain}
\mathcal{F}_s = \bigl[(f_1, \boldsymbol{\theta}_1),\, (f_2, \boldsymbol{\theta}_2),\, \ldots,\, (f_{N_s}, \boldsymbol{\theta}_{N_s})\bigr], \quad s \in \mathcal{S}
\end{equation}
where $\mathcal{S} = \{\text{vocals}, \text{bass}, \text{drums}, \text{other}\}$, $f_i$ is a specific effect name (\eg{}, \texttt{sox\_compand}), $\boldsymbol{\theta}_i$ its parameters (\eg{}, threshold, ratio), and $N_s$ varies per stem.
We concatenate the 4 per-stem FX chains into the full chain $\mathcal{F} = \{\mathcal{F}_s\}_{s \in \mathcal{S}}$, which is serialized into a single token sequence $\mathbf{y}$ for autoregressive prediction (Section~\ref{sec:tokenization}).

\subsection{\bsfilm{} Encoder}\label{sec:encoder}

The encoder maps the 8-channel input $\mathbf{x}$ to a mixing style embedding $\mathbf{e} \in \mathbb{R}^{512}$, following a multi-band CNN design inspired by Kim~\etal{}~\cite{kim2021multiband} and adapted for representation learning with FiLM conditioning.
Each of the 8 channels is transformed into a log-mel spectrogram with $n_{\mathrm{fft}} = 2048$, hop $= 512$, and $n_{\mathrm{mels}} = 80$, yielding shape $(B, 8, 80, T')$.
The frequency axis is divided into overlapping sub-bands of 20 mel bins with 10-bin overlap. Each sub-band is processed by an independent 2-layer CNN with $7{\times}7$ kernels, 32 $\to$ 64 channels, batch normalization, and max pooling, producing per-band feature maps.

FiLM~\cite{perez2018film} injects 64 hand-crafted mixing features into each CNN layer, as detailed in Table~\ref{tab:features}.
These features are adapted from the Diff-MST~\cite{vanka2024diffmst} mixing feature set to our four-stem setting, computing dynamics, spectral, stereo, and inter-stem descriptors per stem.
A 2-layer MLP generates scale $\boldsymbol{\gamma}$ and shift $\boldsymbol{\beta}$ parameters:
\begin{equation}\label{eq:film}
\mathbf{x}' = \boldsymbol{\gamma} \odot \mathbf{x} + \boldsymbol{\beta}, \quad (\boldsymbol{\gamma}, \boldsymbol{\beta}) = \mathrm{MLP}(\mathbf{f})
\end{equation}
The per-band feature maps are concatenated along the channel axis, flattened across frequency, and temporally pooled via learned attention weights to produce the 512-dimensional embedding (1.79M parameters total).
Independent per-band CNNs allow each sub-band to specialize in detecting frequency-specific processing such as band-limited EQ, compression, or de-essing.

\begin{table}[t]
\centering
\caption{Hand-crafted mixing features for FiLM conditioning (64 total).}
\label{tab:features}
\footnotesize
\setlength{\tabcolsep}{2.5pt}
\renewcommand{\arraystretch}{1.15}
\begin{tabular}{@{}llrr@{}}
\toprule
\textbf{Category} & \textbf{Feature (formula)} & \textbf{/Stem} & \textbf{Total} \\
\midrule
\multirow{3}{*}{Dynamics}
  & RMS: $\sqrt{\frac{1}{T}\sum x_c^2}$ ($\times$2 ch.) & \multirow{3}{*}{6} & \multirow{3}{*}{24} \\
  & Crest: $20\log_{10}\!\bigl(\frac{\max|x_c|}{\mathrm{RMS}_c}\bigr)$ ($\times$2 ch.) & & \\
  & K-weighted loudness ($\times$2 ch.): & & \\
  & $-0.691 + 10\log_{10}\!\bigl(\sum_c G_c \cdot \mathrm{RMS}_c^2\bigr)$ & & \\
\midrule
\multirow{3}{*}{Spectral}
  & Band energy: $\bar{E}_{\mathrm{low}}, \bar{E}_{\mathrm{mid}}, \bar{E}_{\mathrm{high}}$ & \multirow{3}{*}{5} & \multirow{3}{*}{20} \\
  & Tilt: $\rho(\text{bin index},\, \bar{E}_{\mathrm{mel}})$ & & \\
  & Flatness: $\exp(\overline{\log S}) \,/\, \bar{S}$ & & \\
\midrule
\multirow{3}{*}{Stereo}
  & ILD: $20\log_{10}(\mathrm{RMS}_L / \mathrm{RMS}_R)$ & \multirow{3}{*}{3} & \multirow{3}{*}{12} \\
  & Correlation: $\sum \tilde{L}\tilde{R} \,/\, \|\tilde{L}\|\|\tilde{R}\|$ & & \\
  & Mid-side ratio: $E_{\mathrm{side}} / E_{\mathrm{mid}}$ & & \\
\midrule
Rel.\ loud. & $L_{\mathrm{stem}} - L_{\mathrm{mix}}$ & 1 & 4 \\[2pt]
Masking & $\overline{\sigma\!\bigl(-(E_s - \max_{j \neq s} E_j)\bigr)}$ & 1 & 4 \\
\bottomrule
\end{tabular}
\renewcommand{\arraystretch}{1.0}
\vspace{2pt}
\par\noindent{\scriptsize
$x_c$: waveform of channel $c$;\;
$S$: mel spectrogram;\;
$\bar{E}$: mean energy over time and channels;\;
$\tilde{L}, \tilde{R}$: zero-mean left/right channels;\;
$E_{\mathrm{mid}} = \overline{((L{+}R)/2)^2}$,\;
$E_{\mathrm{side}} = \overline{((L{-}R)/2)^2}$;\;
$\sigma$: sigmoid;\;
$E_s$: mel energy of stem $s$;\;
$G_c$: channel weighting per ITU-R BS.1770.
}
\end{table}

\subsection{FX Chain Tokenization}\label{sec:tokenization}
{\lsstyle
We serialize variable-length, multi-stem FX chains into flat token sequences.
The vocabulary of 358 tokens comprises 4 special tokens, 4 stem tokens, 85 effect name tokens, 164 parameter name tokens, and 101 quantized value bins.
An example sequence is:
\par}
\smallskip
{\footnotesize
\begin{verbatim}
[BOS] vocals sox_gain sox_gain_gain_db value_55
  sox_compand sox_compand_threshold_db value_30
  sox_compand_ratio value_50 ... [SEP] [EOS]
\end{verbatim}
}
\smallskip
\noindent Continuous parameter values are normalized to $[0,1]$ and quantized to 101 bins.
Frequency-domain parameters such as filter cutoffs use log-scale normalization to match perceptual sensitivity,
$\hat{v} = (\log v - \log v_{\min}) / (\log v_{\max} - \log v_{\min})$,
while all other parameters such as gain, ratio, and attack/release times use linear normalization,
$\hat{v} = (v - v_{\min}) / (v_{\max} - v_{\min})$.
The vocabulary is extensible; adding a new effect requires only a name token, parameter tokens, and normalization ranges.

\subsection{FX Chain Generator}\label{sec:generator}

Both $\mathbf{x}^{\mathrm{orig}}$ and $\mathbf{x}^{\mathrm{aug}}$ pass through the shared \bsfilm{} Encoder to produce 2 embeddings $\mathbf{e}_{\mathrm{orig}}$ and $\mathbf{e}_{\mathrm{aug}}$.
Each embedding is independently projected via a shared projection head consisting of LayerNorm, Linear, ReLU, and Linear layers, producing conditioning vectors $\mathbf{m}_1, \mathbf{m}_2 \in \mathbb{R}^{512}$ that serve as key-value entries for the decoder's cross-attention.

Our paired-embedding design is inspired by AFx-Rep~\cite{steinmetz2024stito}, which encodes both original and augmented audio to learn effect representations from their paired embeddings.
We adopt this paired-input paradigm but replace the classification head with a 6-layer Transformer decoder with $d_{\mathrm{model}} = 512$, 8 heads, and FFN dimension 2048. The decoder attends to $[\mathbf{m}_1, \mathbf{m}_2]$ via cross-attention and projects to the full vocabulary via a linear head, predicting all token types in a unified output space.
Cross-attention separates audio conditioning from the token sequence, allowing the decoder to query the style information independently at each layer.
An alternative is to prepend $\mathbf{m}_1, \mathbf{m}_2$ as prefix tokens. We leave the comparison of conditioning strategies to future work.

\subsection{Training Objective}\label{sec:training}

The ground-truth FX chain $\mathcal{F}$ is serialized into a token sequence $\mathbf{y} = (y_1, \ldots, y_L)$.
The decoder is trained with teacher forcing, minimizing cross-entropy at each step.
\begin{equation}\label{eq:loss}
\mathcal{L} = \mathcal{L}_{\mathrm{CE}} = -\sum_{t} \log p(y_t \mid y_{<t},\, \mathbf{m}_1, \mathbf{m}_2)
\end{equation}
where $y_{<t}$ denotes the ground-truth prefix.

The decoder applies a causal mask so that each position attends only to preceding tokens and the conditioning vectors $\mathbf{m}_1, \mathbf{m}_2$.
Training uses AdamW with learning rate $3{\times}10^{-4}$, weight decay 0.01, linear warmup over 1000 steps, cosine decay, batch size 64, and gradient clipping at norm 1.0.
We train jointly for 60 epochs, approximately 56 GPU hours, on a single NVIDIA RTX 5090 GPU using 10-second audio clips.
The full model comprises the \bsfilm{} Encoder with 1.79M parameters and the 6-layer Transformer decoder with 26.2M parameters, totaling 28.0M parameters.

\subsection{Inference}\label{sec:inference}

At inference time, given a pair of audio inputs (original and target), the encoder produces conditioning vectors $\mathbf{m}_1, \mathbf{m}_2$ that summarize the mixing style difference.
The decoder then autoregressively generates the full token sequence via greedy decoding, predicting effect names, parameter names, and quantized values in a single pass.
Unlike generative models that aim to produce a range of plausible variants, our goal is to recover the chain that comes as close as possible to the ground truth, so we use greedy decoding instead of sampling.
The predicted tokens are parsed back into a structured FX chain specification, mapping each stem to its ordered effect sequence with associated parameter values.

\subsection{\sepaug{} Pipeline}\label{sec:sepaug}
We propose the \sepaug{} Pipeline, which applies source separation to a large mixture dataset and then augments the resulting pseudo-stems with known FX chains to generate paired training data at scale, lifting the multitrack bottleneck described in Section~\ref{sec:intro}.

The pipeline has three stages.
First, a pretrained source separation model splits each song into four pseudo-stems (vocals, bass, drums, other).
Second, \emph{cross-song stem mixing} draws each stem in a training example from a different song, decorrelating musical content from mixing characteristics and preventing the encoder from relying on song-level cues such as key or tempo.
Third, each stem is independently augmented with random effects from \multiafx{}, with a macro-category constraint preventing consecutive effects from the same category.
The result is paired data with complete parameter metadata for supervised training.
See Section~\ref{sec:setup} for more details.

\noindent\textbf{\multiafx{} Toolkit.}
We introduce \multiafx{}, a Python library wrapper unifying seven backends~\cite{sox2005,jordal2020audiomentations,yang2022torchaudio,virtanen2020scipy,mcfee2015librosa,harris2020numpy,steinmetz2021pyloudnorm} behind a consistent interface with 85 effects across 12 macro-categories.
The library supports chain generation with macro-category constraints, parameter range definitions with linear/log normalization, and fault-tolerant batch processing.

\begin{figure*}[t]
\centering
\includegraphics[width=0.95\textwidth]{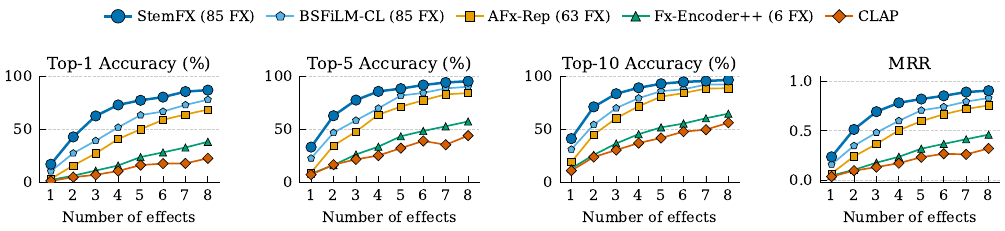}
\vspace{-8pt}
\caption{Mixing style retrieval vs.\ number of applied effects: Top-1, Top-5, Top-10 accuracy (\%) and Mean Reciprocal Rank (MRR).
\stemfx{} outperforms all baselines across all chain lengths and metrics, including BSFiLM-CL (same architecture and data, contrastive training). Numbers in parentheses denote the number of effects each model was trained with.}
\label{fig:retrieval}
\end{figure*}

\section{Experiments}\label{sec:experiments}

\subsection{Setup}\label{sec:setup}

\noindent\textbf{Training Data.}
We apply the \sepaug{} Pipeline to the open FMA dataset~\cite{defferrard2017fma} using SCNet~\cite{li2024scnet} for source separation, producing about 105K songs with 4 pseudo-stems each.
Each song is augmented with 1--10 random effects per stem from \multiafx{}.
Because FMA contains predominantly mono or stereo mixtures rather than true multitracks, many songs lack meaningful energy in one or more separated stems.
We measure integrated loudness for every stem and exclude those below $-40$~dB.
Cross-song stem mixing does not prevent the encoder from observing inter-stem mixing relationships: the encoder still receives all four stems jointly and can learn how their relative levels, spectral balance, and spatial placement interact, regardless of whether they originate from the same song.
However, cross-song mixing may produce disharmonious combinations, such as stems in different keys or tempos. We leave the investigation of musical coherence effects on learned representations to future work.

\noindent\textbf{Evaluation Data.}
For all experiments we use MUSDB18~\cite{rafii2017musdb}, which provides professionally mixed multitracks not seen during training by any model.
To avoid evaluating on tracks with near-silent stems, we select the most balanced 10-second window per track by computing per-stem RMS energy and requiring every stem to contribute at least 10\% of total energy.
Stems are then FX-normalized following Martinez-Ramirez~\etal{}~\cite{martinezramirez2022automix} to a common baseline before augmentation with \texttt{pedalboard}~\cite{sobot2021pedalboard} effects at chain lengths of 1--8.
\texttt{pedalboard} is not included in \multiafx{} and was not used by any of the baseline methods during training, ensuring that all systems are evaluated on out-of-domain audio effects.
FX normalization is used only to construct evaluation pairs. It is not applied during training, where the original and augmented views of a pseudo-stem already differ by exactly the chain to be predicted, so normalizing to a common baseline is unnecessary.

\subsection{Mixing Style Retrieval}\label{sec:retrieval}

We evaluate whether the learned embeddings capture mixing style rather than musical content.
Following the protocol of Fx-Encoder++~\cite{koo2023fxencoder,yeh2025fxencoder}, both the query and the candidate pool consist of augmented stem sets. The task is to retrieve, from a pool of $N = 500$ candidates, the one that shares the same FX chain as the query but contains different musical content. Retrieval is performed by cosine similarity between embeddings.

\noindent\textbf{Baselines.}
We compare against three external encoders using released pretrained weights.
\textbf{AFx-Rep}~\cite{steinmetz2024stito} uses PANNs with mid/side input and an effect classification pretext task.
\textbf{Fx-Encoder++}~\cite{yeh2025fxencoder} is a contrastive stereo encoder producing 2048-d embeddings.
\textbf{CLAP}~\cite{elizalde2023clap} uses HTSAT-tiny with audio--text contrastive learning and 512-d embeddings.
These baselines differ in training data and augmentation, so performance gaps cannot be attributed to the objective alone.
To control for this, we additionally train \textbf{BSFiLM-CL}, the \bsfilm{} Encoder trained with NT-Xent contrastive learning~\cite{chen2020simclr} on the same \sepaug{} data. Following the Fx-Encoder++ paradigm, positive pairs share the same FX chain applied to different content, but we omit the instrument-level query mechanism since \bsfilm{} processes separated stems directly.
This isolates the training objective as the sole variable between \stemfx{} and BSFiLM-CL.
We do not include a comparable controlled experiment for AFx-Rep's classification objective, as the \sepaug{} Pipeline generates over 100K distinct FX chain variants, making closed-set classification impractical at this scale.
Figure~\ref{fig:retrieval} reports Top-1, Top-5, Top-10 accuracy and Mean Reciprocal Rank (MRR).
\stemfx{} outperforms all baselines across all FX counts and all metrics.
At 1~effect, the task is most challenging because subtle parameter differences must be distinguished.
\stemfx{} achieves 16.8\% Top-1 and 0.235 MRR versus 3.0\% Top-1 and 0.067 MRR for AFx-Rep.
At 8~effects, \stemfx{} reaches 86.8\% Top-1 and 0.903 MRR versus 68.4\% and 0.754 for AFx-Rep and 38.0\% and 0.459 for Fx-Encoder++.
The BSFiLM-CL comparison reveals two insights.
First, BSFiLM-CL reaches 77.8\% Top-1 at 8 effects, 9.0 percentage points below \stemfx{} (86.8\%), demonstrating that under matched architecture and data conditions, the FX chain prediction objective produces stronger mixing representations than contrastive learning.
Second, BSFiLM-CL substantially outperforms Fx-Encoder++ despite sharing the same contrastive training paradigm, indicating that the \bsfilm{} Encoder combined with the \sepaug{} Pipeline yields a stronger foundation than Fx-Encoder++'s architecture and training data.

\begin{figure*}[t]
\centering
\includegraphics[width=0.95\textwidth]{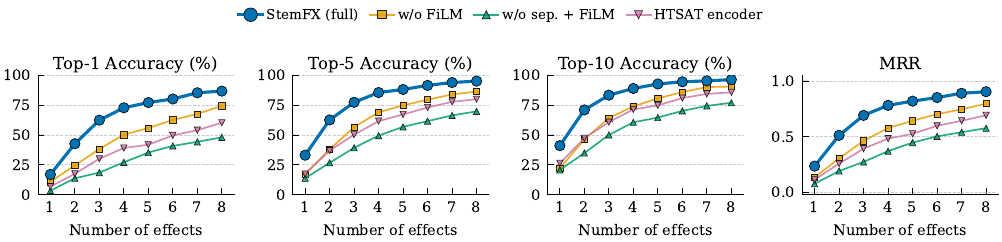}
\vspace{-8pt}
\caption{Ablation study: retrieval performance vs.\ number of applied effects for each model variant.
\stemfx{} (full) is the proposed model; each variant removes one component.}
\label{fig:ablation}
\end{figure*}

\begin{figure*}[t]
\centering
\includegraphics[width=0.95\textwidth]{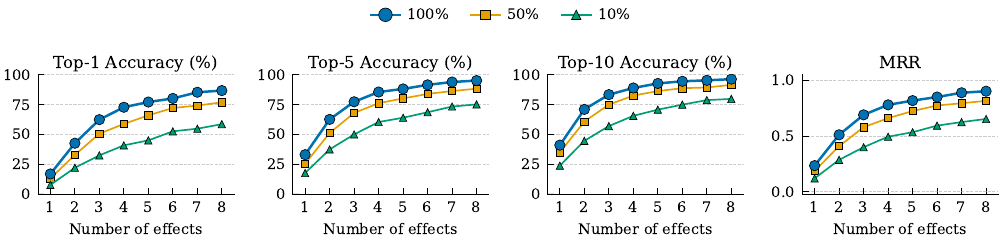}
\caption{Dataset scale analysis: retrieval vs.\ number of applied effects at training set sizes of 10\%, 50\%, and 100\% of 105K songs. The \sepaug{} Pipeline enables training at a scale orders of magnitude beyond existing paired datasets.}
\label{fig:scale}
\end{figure*}

\begin{table}[t]
\centering
\caption{Paired mixing style transfer results.
MUSHRA score is the mean rating from 20 listeners on a 0--100 similarity scale.
\stemfx{} forced and both ITO methods share the same fixed signal chain.}
\label{tab:inference}
\footnotesize
\setlength{\tabcolsep}{2pt}
\begin{tabular}{@{}l@{\hspace{4pt}}rrr r@{}}
\toprule
& Synth. & \multicolumn{2}{c}{Real Mix} & \\
\cmidrule(lr){2-2} \cmidrule(lr){3-4}
& MRSTFT & MRSTFT & MUSHRA & \\[-2pt]
Method & $\downarrow$ & $\downarrow$ & $\uparrow$ & Time \\
\midrule
\textbf{\stemfx{}}    & \textbf{2.35} & \textbf{1.44} & \textbf{60.6} & \textbf{0.24s} \\
\stemfx{} forced      & 3.12 & 2.10 & 25.3 & 1.91s \\
ITO + AFx-Rep         & 3.71 & 2.53 & 30.6 & 1033s \\
ITO + BSFiLM          & 2.89 & 2.06 & 19.9 & 1717s \\
\midrule
Target mix            & 0.00 & 0.00 & 96.6 & --- \\
FX-normalized         & 3.38 & 1.49 & 54.9 & --- \\
\bottomrule
\end{tabular}
\end{table}

\subsection{Paired Mixing Style Transfer}\label{sec:inference_comparison}

Given a paired original and target, \stemfx{} can transfer the target's mixing style by predicting and executing the FX chain that bridges them.
We evaluate on two test sets.
The \emph{Synthetic} set contains 20 MUSDB18 pairs where targets are augmented with 8 \texttt{pedalboard} effects per stem, out-of-domain from any training data, stress-testing generalization across a wide range of devices and parameter values.
The \emph{Real Mix} set takes FX-normalized stems as input and the original professionally mixed MUSDB18 tracks as the target, testing whether methods can recover mixing decisions in real-world scenarios.
We conduct a MUSHRA-style listening test with 20 listeners with music production experience on the Real Mix set only. Listeners rate how similar each condition sounds to the target mix on a 0--100 scale in a blind, randomized setting. Each song includes the four method outputs alongside two hidden references: the target mix itself as the high anchor and the FX-normalized input as the low anchor. The Synthetic set is excluded from listening because up to eight randomly combined effects produce unnaturally dense processing, making perceptual quality judgments difficult.
We measure spectral fidelity via multi-resolution STFT loss, MRSTFT, and inference time. Table~\ref{tab:inference} reports the results.

We compare four methods.
\textbf{\stemfx{}} uses autoregressive generation with default settings, predicting both which effects to apply and their parameters.
All three fixed-chain methods operate on the same universal signal chain covering gain staging, highpass/lowpass filters, four parametric EQ bands, compression, distortion, chorus, delay, and reverb.
Each device can be effectively bypassed with neutral parameter values, so the chain spans styles from minimal processing to complex multi-effect configurations.
\textbf{\stemfx{} forced} predicts only value tokens for this fixed chain, with structural tokens injected from the chain template.
The two ITO baselines follow the inference-time optimization approach of ST-ITO~\cite{steinmetz2024stito}, optimizing the same universal chain via CMA-ES with population 64, $\sigma_0{=}0.3$, and 25 iterations. Each of the four stems is processed independently.
\textbf{ITO $+$ AFx-Rep} uses AFx-Rep~\cite{steinmetz2024stito} cosine similarity as the optimization objective, while \textbf{ITO $+$ BSFiLM} substitutes the \bsfilm{} Encoder.

\noindent\textbf{Synthetic set.}
\stemfx{} achieves the best MRSTFT of 2.35 at 0.24s per example, over 4000 times faster than iterative optimization.
The forced variant enables a controlled comparison with ITO, as all three share the same signal chain and parameter space, isolating feedforward prediction from iterative search.
Neither ITO variant beats \stemfx{}, indicating that selecting the right effects with proper order matters more than perfectly optimizing parameters of a predetermined chain.

\noindent\textbf{Real Mix set.}
On professional mixes, the advantage of free generation becomes even more pronounced.
\stemfx{} achieves the best MRSTFT of 1.44 and the highest MUSHRA score of 60.6, the only method that surpasses the FX-normalized low anchor of 54.9.
The high anchor scores 96.6, confirming that listeners can reliably identify the target mix.
All fixed-chain methods score 19.9 to 30.6. Listening to low-scoring cases confirms that these methods frequently apply audible distortion or modulation where the target mix uses neither.
The universal chain exposes effects that many target mixes do not use, and driving every such effect to neutral parameters is difficult. ITO must jointly optimize all parameters in the chain without a prior for which effects to bypass, and \stemfx{} forced likely suffers from distribution mismatch since the training data rarely contains neutral-parameter instances.
\stemfx{} avoids both failure modes by selecting only the effects that match the target, a fundamental advantage of the autoregressive formulation which chooses \emph{which} effects to apply, not only \emph{how} to parameterize a fixed set.
We also demonstrate unpaired style transfer on the demo page.\footnote{\url{https://barry-mir.github.io/stemfx-demo/}}

\subsection{Ablation Study}\label{sec:ablation}

Figure~\ref{fig:ablation} isolates the contribution of each \stemfx{} component using three ablation variants.
Removing FiLM conditioning reduces Top-1 from 86.8\% to 74.4\% at 8 effects, confirming that hand-crafted mixing features provide useful inductive bias.
Replacing source-separated stems with stereo mixture input causes the largest degradation, dropping Top-1 to 48.0\%.
This variant also removes FiLM conditioning, since the hand-crafted features in Table~\ref{tab:features} are per-stem and undefined for mixture input.
The gap between w/o FiLM at 74.4\% and w/o sep.\ + FiLM at 48.0\% isolates source separation, showing that per-stem input accounts for 26.4 percentage points of improvement even without FiLM.
Replacing the \bsfilm{} Encoder with HTSAT-tiny~\cite{chen2022htsat} drops Top-1 to 60.2\%, showing that the stem-aware multi-band CNN design outperforms general audio transformers for mixing style representation.

\subsection{Dataset Scale Analysis}\label{sec:scale}

Figure~\ref{fig:scale} shows retrieval accuracy as a function of training set size (10\%, 50\%, 100\% of 105K songs).
Performance improves consistently from 58.6\% at 10\% size to 77.0\% at 50\% size and 86.8\% at 100\% size,
demonstrating that the \sepaug{} Pipeline's ability to scale training data translates directly into better representations.
This scaling behavior is not achievable with real multitrack datasets, which are limited to hundreds of songs.

\section{Limitations and Future Work}\label{sec:limitations}

\stemfx{} can only predict effects contained in the FX set it was trained on, and supporting additional effects requires retraining on data that includes them.
\stemfx{} also operates on the four stems produced by current separation systems, a step beyond estimating mixing style from the stereo mixture alone, though extending it to the finer decompositions of a professional session is left to future work.
Pseudo-stems additionally inherit the errors of the separation model, and we do not quantify how these artifacts propagate into the learned representation; a comparison against a model trained on clean multitrack stems would isolate this effect.
Finally, training chains are sampled at random, whereas mix engineers apply structured, genre-dependent processing. Whether the representation captures such conventions beyond random effect compositions remains open.

\section{Conclusion}\label{sec:conclusion}

We presented \stemfx{}, a framework for learning mixing style representations through autoregressive FX chain prediction on source-separated stems, with three contributions:
an end-to-end autoregressive formulation that outperforms contrastive learning under matched conditions, 86.8\% vs.\ 77.8\% Top-1;
the \bsfilm{} Encoder, whose FiLM conditioning and per-stem input contribute 12.4\% and 26.4\%;
and the \sepaug{} Pipeline with the \multiafx{} Toolkit, which scales paired training data to about 105K songs and lifts Top-1 from 58.6\% to 86.8\% as the training set grows.


\section{Acknowledgments}
The work is supported by grants from Google Asia Pacific, the National Science and Technology Council of Taiwan (NSTC 114-2628-E-002-013-MY3), and the Ministry of Education (MOE) of Taiwan (for Taiwan Centers of Excellence in Artificial Intelligence).

\bibliography{references}

\end{document}